\documentclass[12pt]{article}

\usepackage{graphics}
\usepackage{amsmath}
\usepackage{cite}
\usepackage[export]{adjustbox}
\usepackage{color}
\usepackage{graphicx}
\usepackage{dcolumn}
\usepackage{bm}
\usepackage{hyperref}
\usepackage{float}
\usepackage[caption=false]{subfig}
\usepackage{subfloat}
\usepackage{hyperref}
\hypersetup{
colorlinks=true,
linktoc=page,    
linkcolor=blue,
citecolor=blue,
urlcolor=blue,
colorlinks=true,
}

\newcommand{\be}{\begin{equation}}
\newcommand{\ee}{\end{equation}}
\newcommand{\bea}{\begin{eqnarray}}
\newcommand{\eea}{\end{eqnarray}}

\newcommand{\mM}{\mathcal{M}}

\newcommand{\mD}{\mathcal{D}}
\newcommand{\mS}{\mathcal{S}}

\def\/{\frac}
\def\pd{\partial}

\newcommand{\nn}{\nonumber}

\newcommand{\dd}{\,\mathrm{d}}

\newcommand{\bra}[1]{\left<{#1}\right|}
\newcommand{\brat}[2]{\left<{#1}|{#2}\right>}
\newcommand{\ket}[1]{\left|{#1}\right>}
\newcommand{\braket}[2]{\left<{#1}\right|{#2}\left|{#1}\right> }

\begin{document}
\thispagestyle{empty}
\vspace*{.5cm}
\begin{center}

{\bf {\LARGE Replica wormhole \\as a vacuum-to-vacuum transition}}\\

\begin{center}
\vspace{1cm}
{\bf Yang An$^1$ and Peng Cheng$^{*2,3}$\footnotetext[1]{p.cheng.nl@outlook.com(corresponding author)}}\\
 \bigskip \rm
  
\bigskip
$^1$\hspace{.05em}Institute for Theoretical Physics \& Cosmology, Zhejiang University of Technology, 310014 Hangzhou, China\\
$^2$\hspace{.05em}Center for Joint Quantum Studies and Department of Physics, School of Science, 
Tianjin University, 300350 Tianjin, China \\
$^3$\hspace{.05em}Institute for Theoretical Physics, University of Amsterdam, 1090 GL Amsterdam, Netherlands\\

\rm
  \end{center}

\vspace{1cm}
{\bf Abstract}
\end{center}
\begin{quotation}
\noindent

The recent developments related to the black hole information paradox have brought us a confusing object: the replica wormhole. 
We are trying to better understand this object from the viewpoint of the thermo-mixed double and spontaneous symmetry breaking.
In this paper, we show that the replica wormhole can be regarded as a transition between different degenerate vacua, and the corresponding gravitational partition function should be controlled by the manifold of the degenerate vacua.
We also check the wormhole partition function in 2-dimensional Jackiw-Teitelboim gravity and show that the wormhole saddle is indeed controlled by the dimension of the degenerate vacua.
Moreover, it is suggested that the replica wormhole geometries connecting different vacua can be related to the measurement process of soft hair that compares different vacuum configurations.

\end{quotation}

\setcounter{page}{0}
\setcounter{tocdepth}{2}
\setcounter{footnote}{0}
\newpage

\parskip 0.1in
 
\setcounter{page}{2}
\tableofcontents

\section{Introduction}
\label{intro}

The black hole information paradox (BHIP) is a longstanding problem regarding the evolution of black hole systems, and we have seen exciting progress \cite{Penington2019,Almheiri2019,Almheiri2019a,Almheiri2019c,Penington2019a,Almheiri2019b,Almheiri2020} in recent years. 
The Page cure \cite{Page1993,Page2013} was derived from the quantum extremal surface prescription \cite{Ryu:2006bv,Hubeny:2007xt,Lewkowycz2013,Faulkner:2013ana,Barrella2013,Engelhardt:2014gca,Wall:2012uf}, and extra ``island" contribution to the radiation entropy was the key ingredient for deriving the Page cure.
When using the replica trick and Euclidean path integral to derive the island rule, the authors of \cite{Penington2019a,Almheiri2019b} have found \textit{replica wormhole} saddles which play a vital role in getting the Page curve.

Then, the most crucial question to ask is: how should we understand the replica wormhole?
The replica ansatz \cite{Verlinde2020,Verlinde2021} provided a way of understanding the replica wormhole by relating the $n$-fold wormhole and the thermo-mixed double state. In this paper, we will use a similar idea and show that the replica wormhole can be understood as a transition between different degenerate vacua.

Such ideas are well-motivated because it is natural to relate the wormhole saddle that decreases the radiation entropy and measurement process in soft BHIP story \cite{Pasterski2020,Cheng2020,Cheng2021a}. In QFT, the spontaneous symmetry breaking (SSB) of the global symmetry forms superselection sectors of Hilbert space, and no renormalizable operator can move the system between different sectors.
However, as told by Hawking-Perry-Strominger (HPS) \cite{Hawking2016,Hawking2016a}, there are infinite degenerate vacua at null boundaries because of the SSB of large gauge symmetry, and no superselection sector is formed (related properties at black hole horizon were analyzed in \cite{Cheng:2022xgm,Cheng:2022xyr}).
The degenerate vacua are labeled by local functions, such as $f(z_A)$ on the celestial sphere.
It was suggested in the soft BHIP \cite{Pasterski2020,Cheng2020,Cheng2021a} that measurement processes that compare different functions $f(z_A)$ can decrease the entropy of a black hole system. 
The Hawking radiation process can increase the entropy by producing new radiations, while the soft hair measurement process can decrease the entropy by comparing different $f$s. 
The competition between the Hawking radiation and measurement process gives rise to the Page curve consistent with unitary evolution \cite{Cheng2020}. 
The argument shown in \cite{Pasterski2020,Cheng2020,Cheng2021a} is similar to the competition between connected and disconnected geometries shown in \cite{Penington2019a,Almheiri2019b}. The disconnected ``Hawking saddles'' increase the entropy of the radiation and the black hole, while the connected ``wormhole saddles'' decrease the corresponding entropy.
The aim of the study is to find a clear connection between the recent developments on Euclidean gravitational path integral and the soft BHIP story.

We explicitly show the relation between the replica wormhole and degenerate vacua in this paper. By a careful study of the thermo-mixed double and wormhole saddle, we show there can be different vacua on different boundaries of the wormhole saddle as opposed to the black hole partition function. Moreover, the wormhole partition function should be proportional to the number of possible vacua and the replica wormhole geometry in calculating the radiation entropy should be regarded as a process that decreases the entropy.
We then check the partition function of the wormhole saddle in 2-dimensional Jackiw-Teitelboim (JT) gravity, which serves as evidence for the proposal.

In this paper, we explore how should we understand replica wormholes. The paper is organized as follows. In section \ref{RWH}, we offer a basic overview of how the replica wormhole arises. Then, section \ref{secTMD} provides the argument on how to treat the replica wormholes as a transition from different degenerate vacua. Section \ref{JT} checks the above proposal in 2-dimensional JT gravity. We summarize the paper and discuss further issues in section \ref{con}. The appendix \ref{EPI} reviews the basic concepts of the Euclidean path integral and black hole thermodynamics.

\section{Replica wormholes}
\label{RWH}

In this section, following \cite{Penington2019a}, we will see how the replica wormhole makes its appearance in the argument of the BHIP. The basics of the Euclidean path integral are reviewed in Appendix \ref{EPI}. We are going to use $\ket{\Psi}$ to denote the pure system composed of a black hole and radiation. Then we can trace out the black hole system to calculate the fine-grained entropy of the radiation.
The state $\ket{\Psi}$ can be expressed as
\be
\ket{\Psi}=\frac{1}{\sqrt{k Z_1}}\sum_{i=1}^k\ket{\psi_i}_{B}\ket{i}_R\,,
\ee
where $\ket{\psi_i}_{B}$ denotes the black hole states, $\ket{i}_R$ denotes the radiation, and $i$ labels the entanglement between the radiation and black hole. The normalization factor is chosen such that under
\be
\brat{\psi_i}{\psi_j}_B=\delta_{ij} Z_1\,,~~~~\brat{i}{j}_R=\delta_{ij}\,.
\ee
the amplitude
\be\sum_\Psi \brat{\Psi}{\Psi}=1\,.
\ee
The density matrix of radiation can be obtained by tracing out the black hole system
\be
\rho_R=\sum_m\brat{\psi_m}{\Psi}\brat{\Psi}{\psi_m}=\frac{1}{k}\sum_{i,j}\ket{i}_R\bra{j }\brat{\psi_j}{\psi_i}_B\label{densitytrace}\,,
\ee
which can be illustrated as
\be
\rho_R=
\begin{matrix}
 \includegraphics[width=3cm]{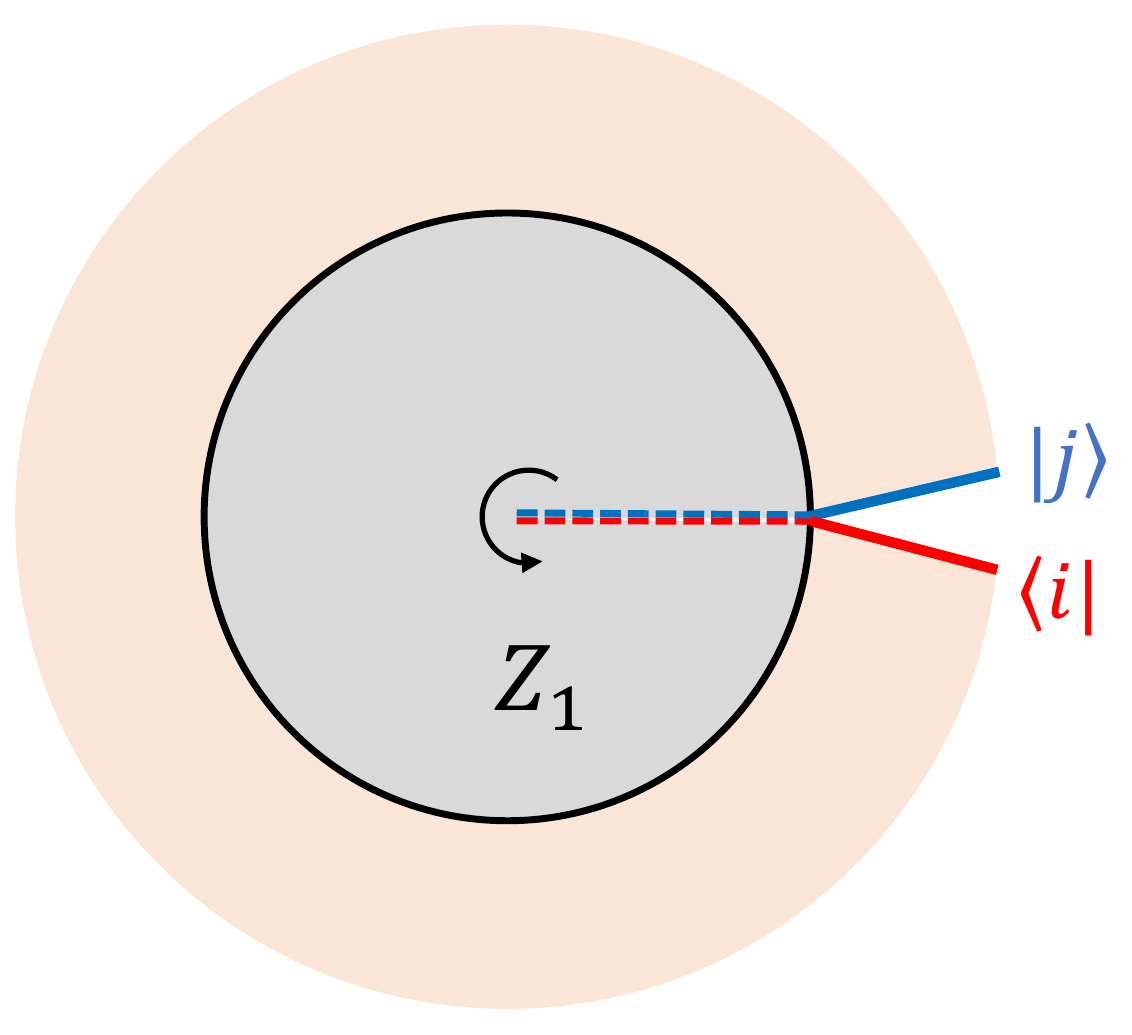}
\end{matrix}\,. \label{single}
\ee
The density matrix is represented in the same way as we described in the previous section.

\begin{figure}
\centering{\includegraphics[width=3.5cm]{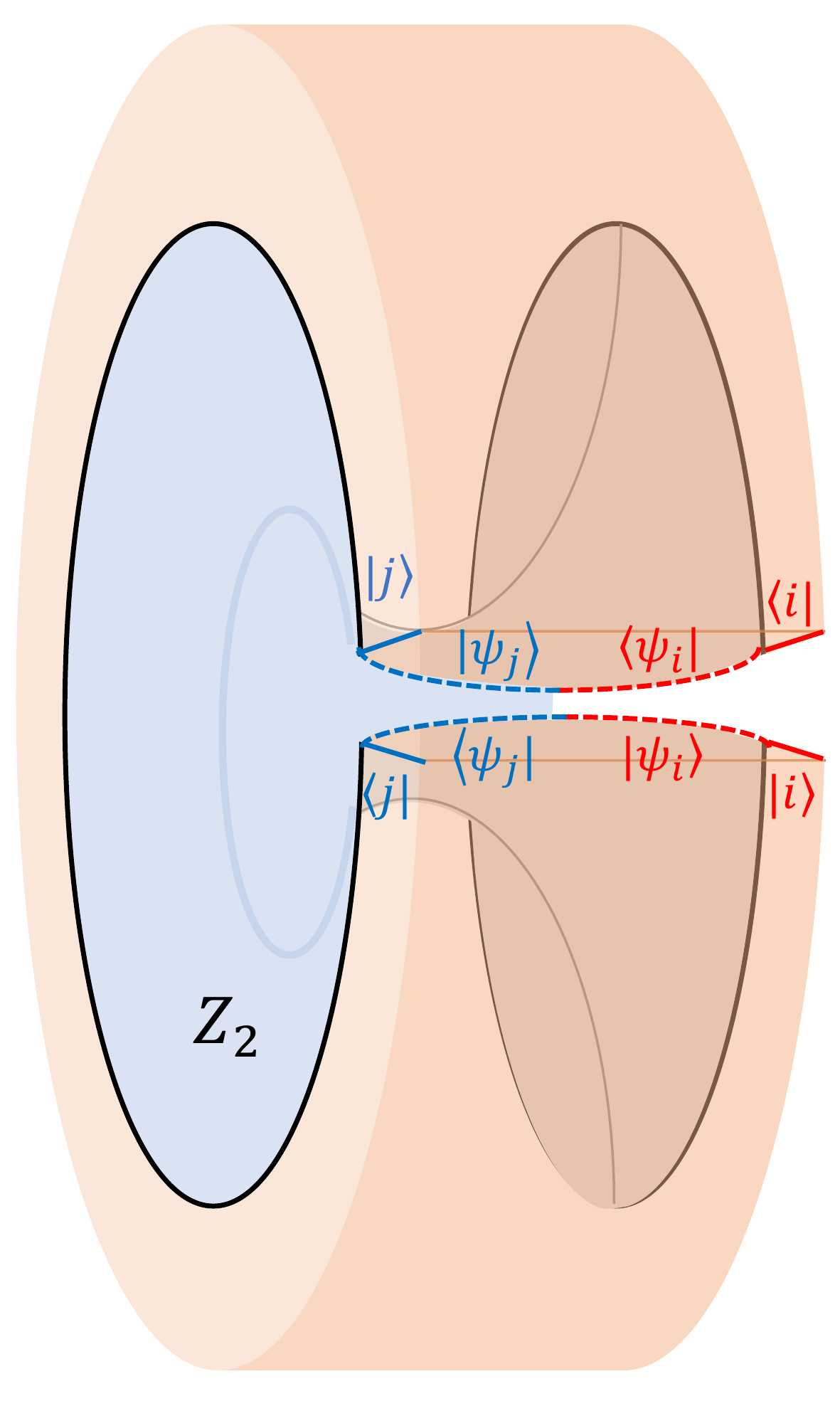}}
\caption{Replica wormhole saddle in calculating the purity of radiation (\ref{purity}). The picture is a diagrammatic representation of $\sum_{i,j}\brat{\psi_i,\psi_j}{\psi_i,\psi_j}$. There is a wormhole geometry
, and the final result is $\sum_{i,j}\brat{\psi_i,\psi_j}{\psi_i,\psi_j}=k^2 Z_2$. Note that the * symbol representing the CPT transformation is not explicitly shown in the figure. }\label{replicaworm}
\end{figure}

\begin{figure}
\centering{
\includegraphics[width=7cm]{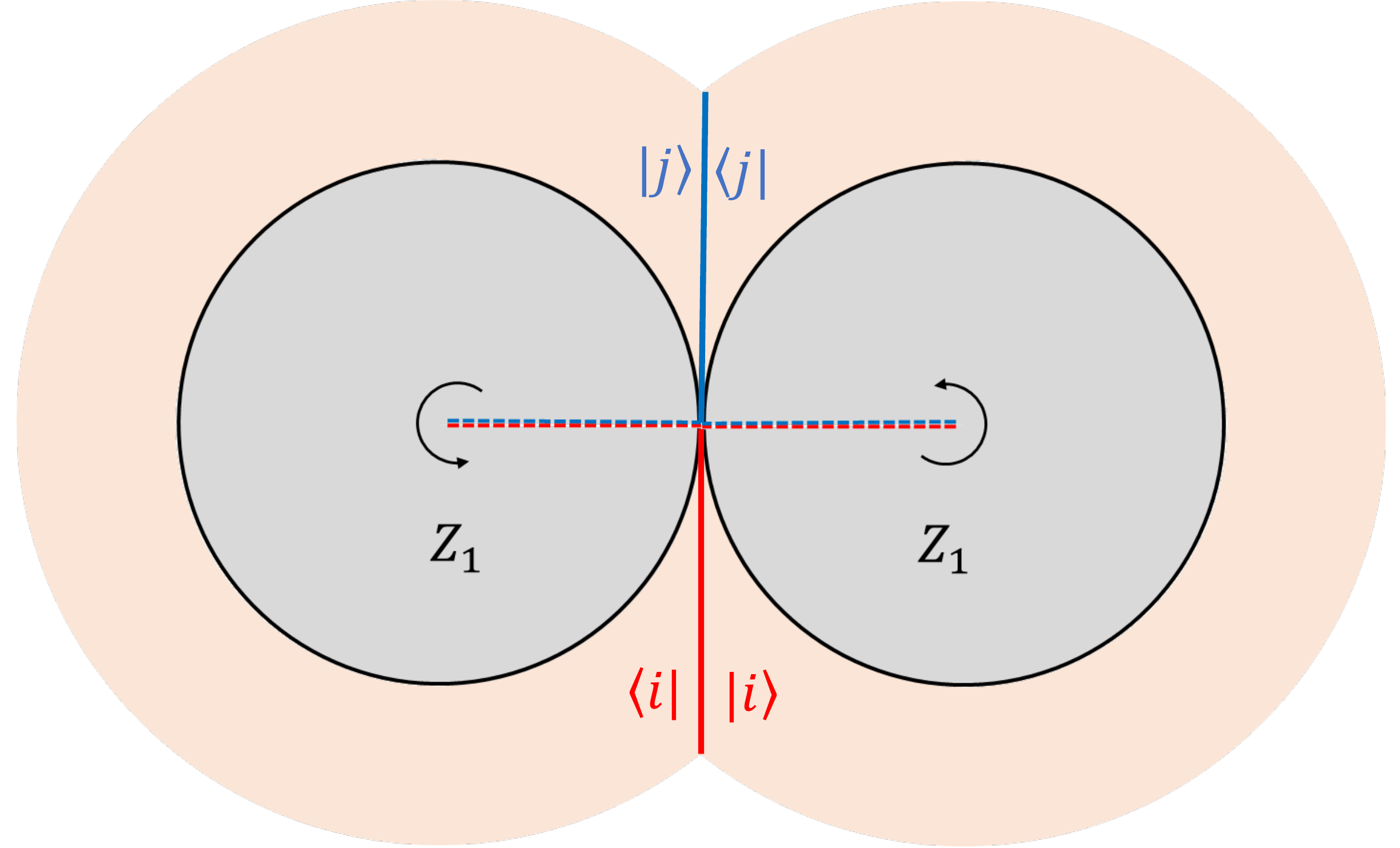}
}
\caption{Hawking saddle in calculating the purity of radiation (\ref{purity}). The corresponding result is $\sum_{i,j}\brat{\psi_i}{\psi_j}\brat{\psi_j}{\psi_i}=k Z_1^2$. }\label{replicahawking}
\end{figure}

Instead of directly using the density matrix to calculate the entropy, one can look at purity tr$[\rho^2_R]$, where we have
\bea
 \text{tr} [\rho_R^2]
&=& \frac{1}{k^2Z_1^2}\sum_l\sum_{i,j}\sum_{m,n}\brat{l}{i}\bra{j}\brat{\psi_j}{\psi_i}\ket{m}\bra{n}\brat{\psi_n}{\psi_m}\ket{l}\nn\\
&=& \frac{1}{k^2Z_1^2}\sum_{i,j}\left[\brat{\psi_i}{\psi_j}\brat{\psi_j}{\psi_i}\right]
+\frac{1}{k^2Z_1^2}\sum_{i,j}\left[\brat{\psi_i,\psi_j}{\psi_i,\psi_j}\right]
\nn\\
&=& \frac{kZ_1^2+k^2 Z_2}{k^2 Z_1^2}\,,\label{purity}
\eea
where $Z_1$ is the geometry described in equation (\ref{single}), and $Z_2$ represents the 2-fold replica wormhole shown in Fig. \ref{replicaworm}.
Now, the purity can be expressed as
\be
\text{tr} [\rho_R^2] =\frac{kZ_1^2+k^2 Z_2}{k^2 Z_1^2} =\frac{1}{k}+\frac{Z_2}{Z_1^2}\,.\label{purity0}
\ee
The calculation shown in (\ref{purity}) is explicitly illustrated in Fig. \ref{replicaworm} and Fig. \ref{replicahawking}.

The 2-fold wormhole shown in Fig. \ref{replicaworm} can be ironed to a flat version
\be
\begin{matrix}
 \includegraphics[width=2.5cm]{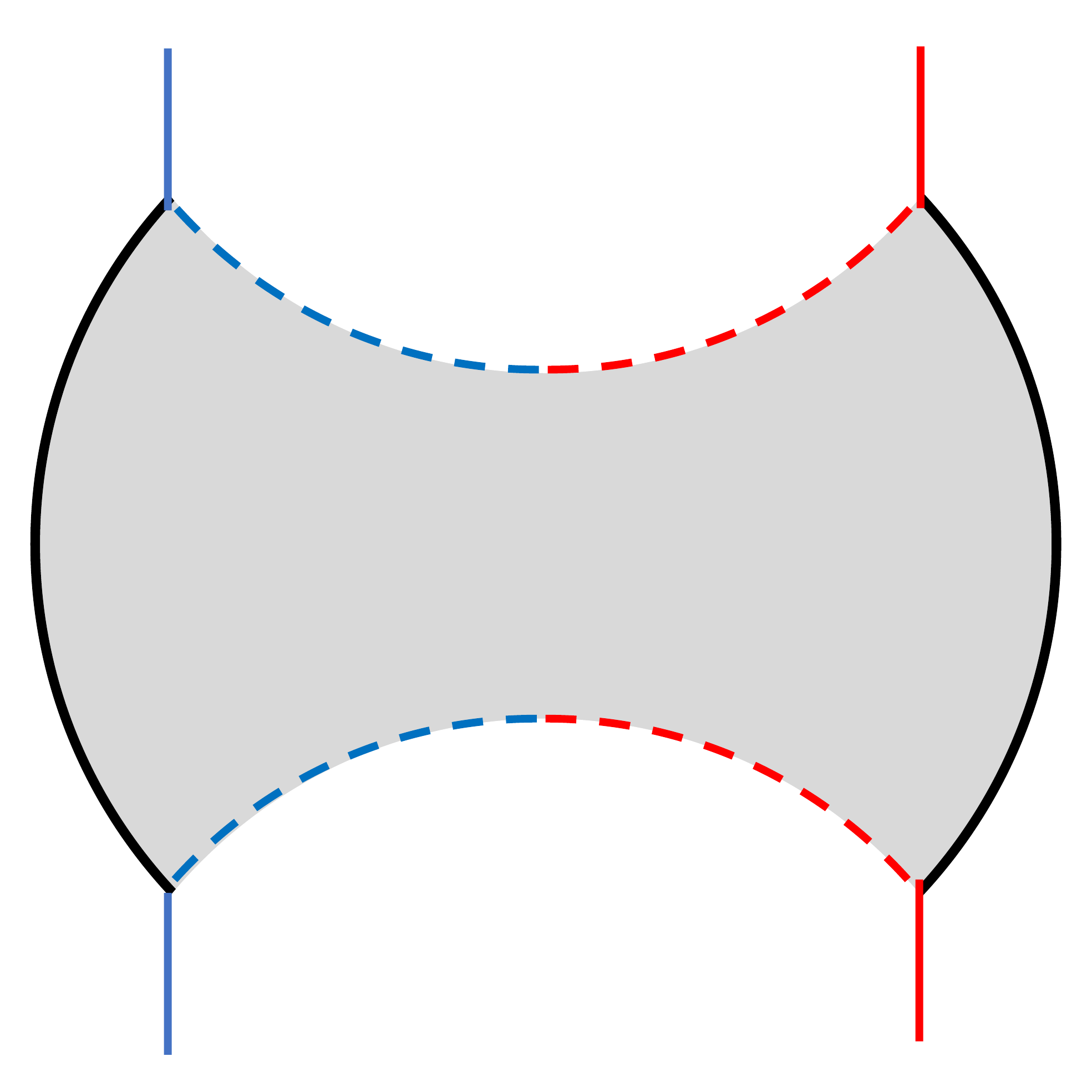}\label{2-fold}
\end{matrix}\,,
\ee
where we only show the gravitational region (grey region) of Fig. \ref{replicaworm}. The wormhole geometry can be recovered by gluing the dashed lines with the same color together in (\ref{2-fold}). The above geometry can be easily extended to $n$-fold replica wormholes, which can be used to calculate the $n$-th Renyi entropy $\text{tr}[\rho_R^n]$.
The von Neumann entropy is essentially determined by $\text{tr}[\rho_R^n]$ when analytically continued to $n=1$.
There are more different saddles for the $n$-th Renyi entropy: fully connected, fully disconnected, and other partially connected saddles. The saddles with more disconnected components dominate at the early stage when $k$ is relatively small. And the connected geometries play a more important role at the late stage of evaporation.
We can add all the different saddles together, and the transitions between different saddles lead to an entropy of radiation that is consistent with the unitarity \cite{Penington2019a,Almheiri2019b}.


\section{Transition between different vacua}
\label{secTMD}

Let us think about what the replica wormhole saddle, shown in Fig. \ref{replicaworm}, means in gravity.
As shown in appendix \ref{EPI}, the Euclidean black hole partition function can be got by tracing the thermofield double (TFD) state:
\be
Z_{BH}=\sum_{\phi}
\begin{matrix}
 \includegraphics[width=2.3cm]{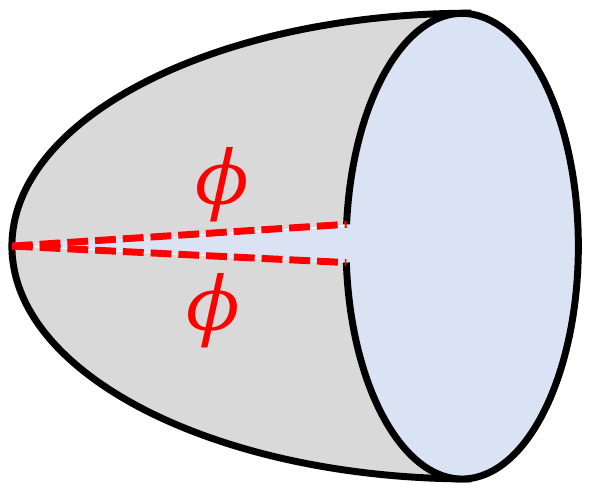}
\end{matrix}=
\begin{matrix}
 \includegraphics[width=2.3cm]{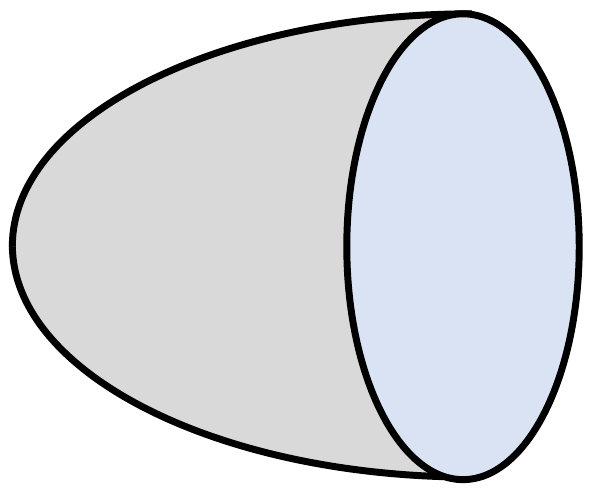}
\end{matrix}\,,
\ee
which is a gravitational generalization of the Rindler partition function obtained from Eq.(\ref{TFD1}).
Looking at the replica wormhole, we find that we have lost our favorite TFD state shown in (\ref{TFD1}). If one insists that the density matrix of the replica wormhole should also be got from tracing over some TFD-like state, we can express the wormhole density matrix $\rho_{\text{WH}}$, with ``WH" representing wormhole, as
\be
\rho_{\text{WH}}= \sum_{\phi_1,\phi_2}
\begin{matrix}
\includegraphics[width=2.5cm]{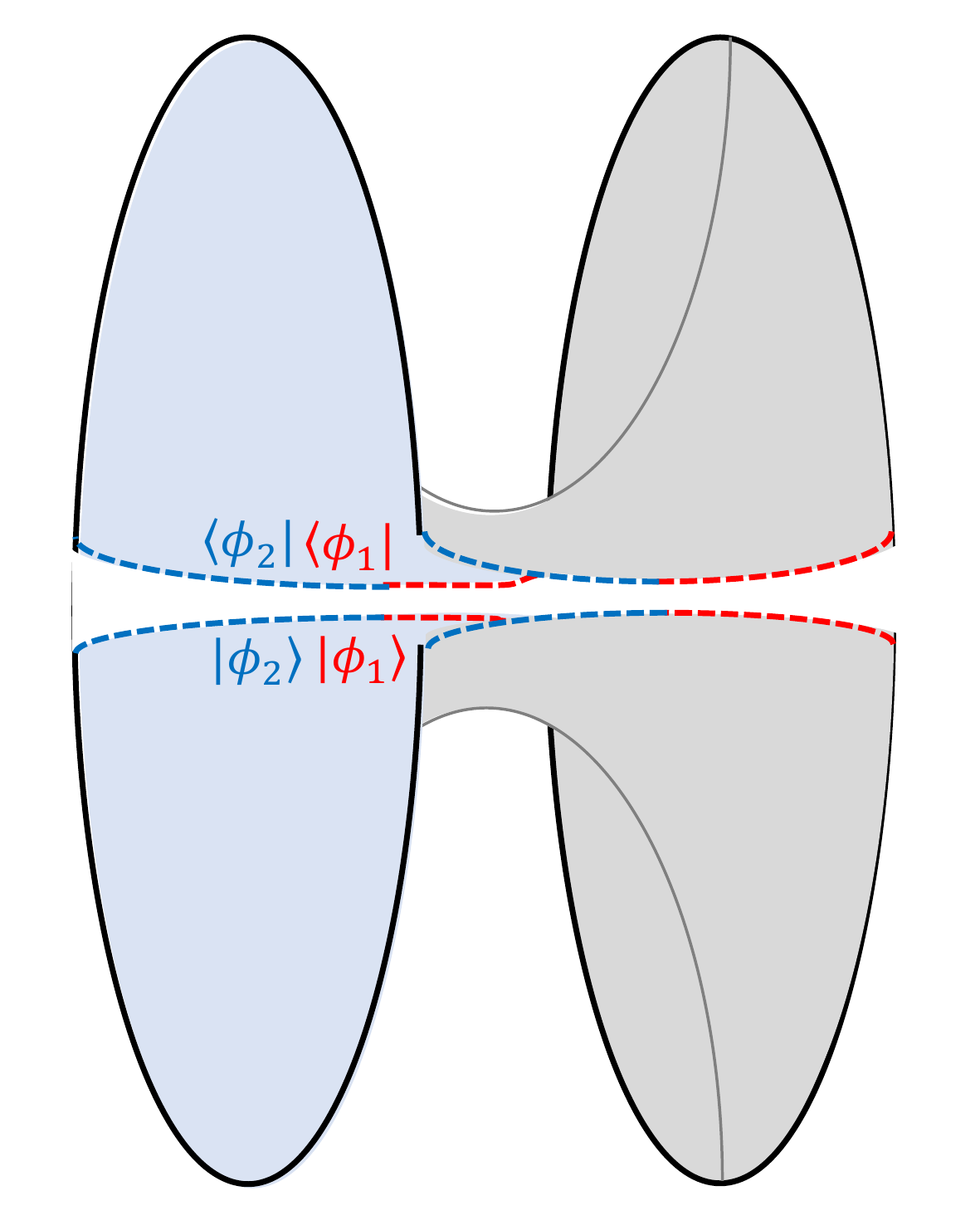}
\end{matrix}=
\begin{matrix}
\includegraphics[width=2.5cm]{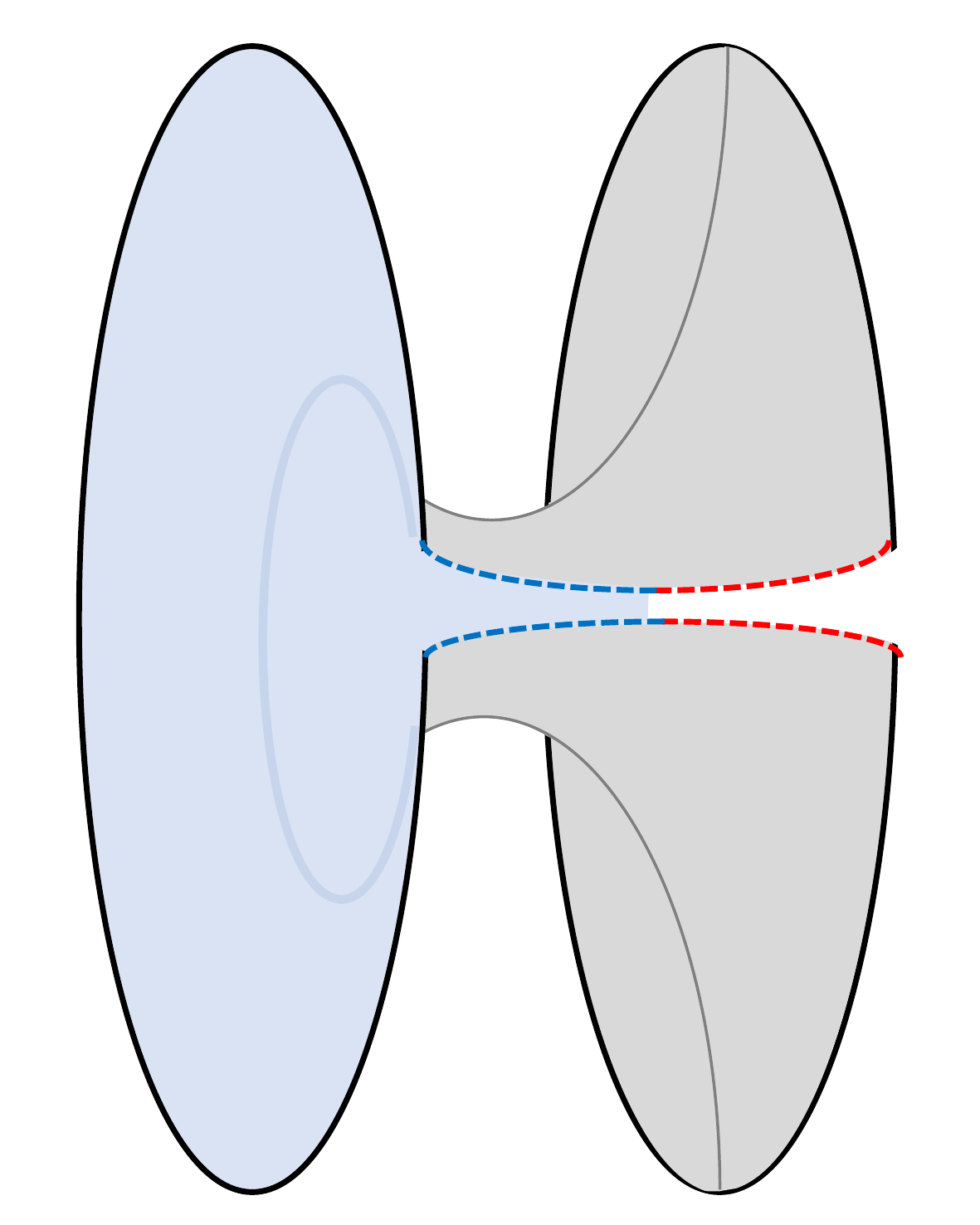}
\end{matrix}\,.\label{TMDdensity}
\ee
Now, assuming the wormhole density matrix is obtained from some reduced density matrix, we have got the density matrix of the \textit{thermo-mixed double} (TMD) \cite{Verlinde2020,Verlinde2021,Campo2019,Anegawa:2020lzw}, which can be illustrated as
\be
\rho_{\text{TMD}}=
\begin{matrix}
\includegraphics[width=2.5cm]{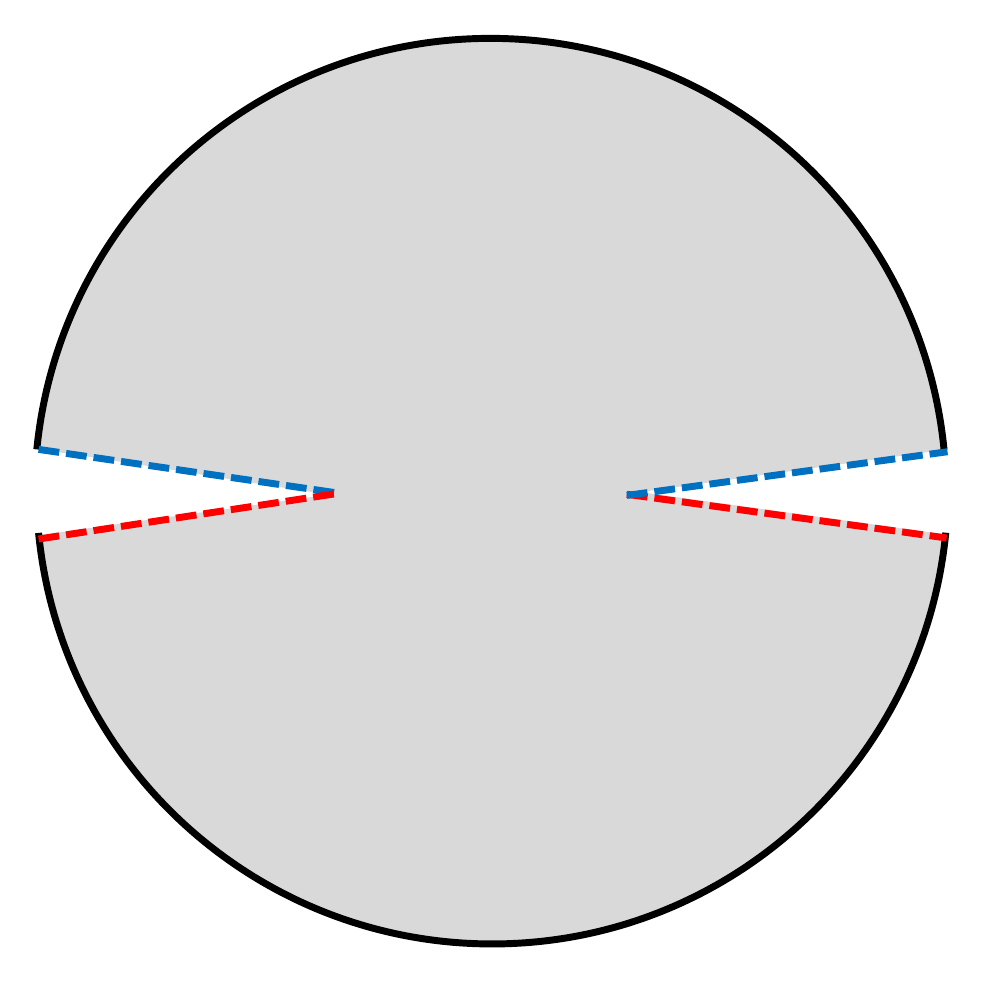}
\end{matrix}\,.\label{TMD0}
\ee
$\rho_{\text{TMD}}$ is the flat version of the up or down part of (\ref{TMDdensity}), and the connected region in the middle of (\ref{TMD0}) can be regarded as the ``island".
The TMD density can be called ``Janus Pacman" to distinguish the Pacman figure shown in (\ref{pacman}). The TMD and TFD are supposed to give out the same density matrix when tracing out the left part $L$
\be
\rho_{BH}=\text{tr}_{L}(\rho_{\text{TMD}})=\text{tr}_{L}(\rho_{\text{TFD}})\,,
\ee
whose von Neumann entropy is the Bekenstein-Hawking entropy $S_{BH}$\footnote{More generally, for $n$-fold wormholes, we have Herman Verlinde's \textit{replica ansatz} \cite{Verlinde2020,Verlinde2021}
\be
\text{tr}(\rho_{\text{TMD}}^n)=\frac{Z_n}{Z_1^n}\,,\label{replicaA}
\ee
with $n$-fold wormhole partition function $Z_n$.}.


Equipped with the TMD interpretation, let us look at what the replica wormhole means now. As discussed in Appendix \ref{EPI}, the TFD state can be defined as a state evolved from the vacuum at $\tau=-\infty$. Here, by the same logic, we can also define the TMD as a state evolved from the vacuum. The vacuum evolved from $\tau=-\infty$ is denoted as $\ket{0_d}$ and the one from $\tau=\infty$ as $\ket{0_u}$.
The subscripts $u$ and $d$ mark different possible vacua.

Those different vacua can be identified by trace operation.
In (\ref{densitytrace}), the trace over the gravity part $\ket{\psi_i}_B$ forces the vacuum to be identical, which can be seen from the closed black circle shown in Eq. (\ref{single}). For the same reason, when the boundaries representing $\ket{0_u}$ and $\ket{0_d}$ are connected with each other, we have a unified vacuum, so putting an extra label on the vacuum state has no influence on the final result because of the trace operation.
In the case of calculating the purity, the situation is exactly shown in Fig. \ref{replicahawking}, where the black line forms a closed circle in the Hawking saddle. So in the Hawking saddle, $\ket{0_u}$ and $\ket{0_d}$ are identified by the trace operation, and thus we have a unified vacuum, which is consistent with our usual understanding of quantum field theory with a given Hamiltonian. However, the situation is completely different for the replica wormhole saddle shown in Fig \ref{replicaworm}.
As can be seen from Fig \ref{replicaworm}, the black lines representing $\ket{0_u}$ and $\ket{0_d}$ form two different circles, and no trace operation forces them to be identical. In principle, there can be different degenerate vacua, and that is the part where the different subscripts on $\ket{0}$ start to play an important role in the follow-up discussions. We leave the physical interpretation of degenerate vacua in the discussion section.

Now, we come to the key observation regarding the meaning of the replica wormhole.
For the flat case corresponding to the Minkowski and Rindler spacetime, the TMD density matrix can be represented as
\bea
\bra{\psi_1}\rho_{\text{TMD}}\ket{\psi_2} &=& \sum_\phi\bra{\phi}\brat{\psi_1}{0_d}\brat{0_u}{\psi_2}\ket{\phi}=
\begin{matrix}
\includegraphics[width=2.8cm]{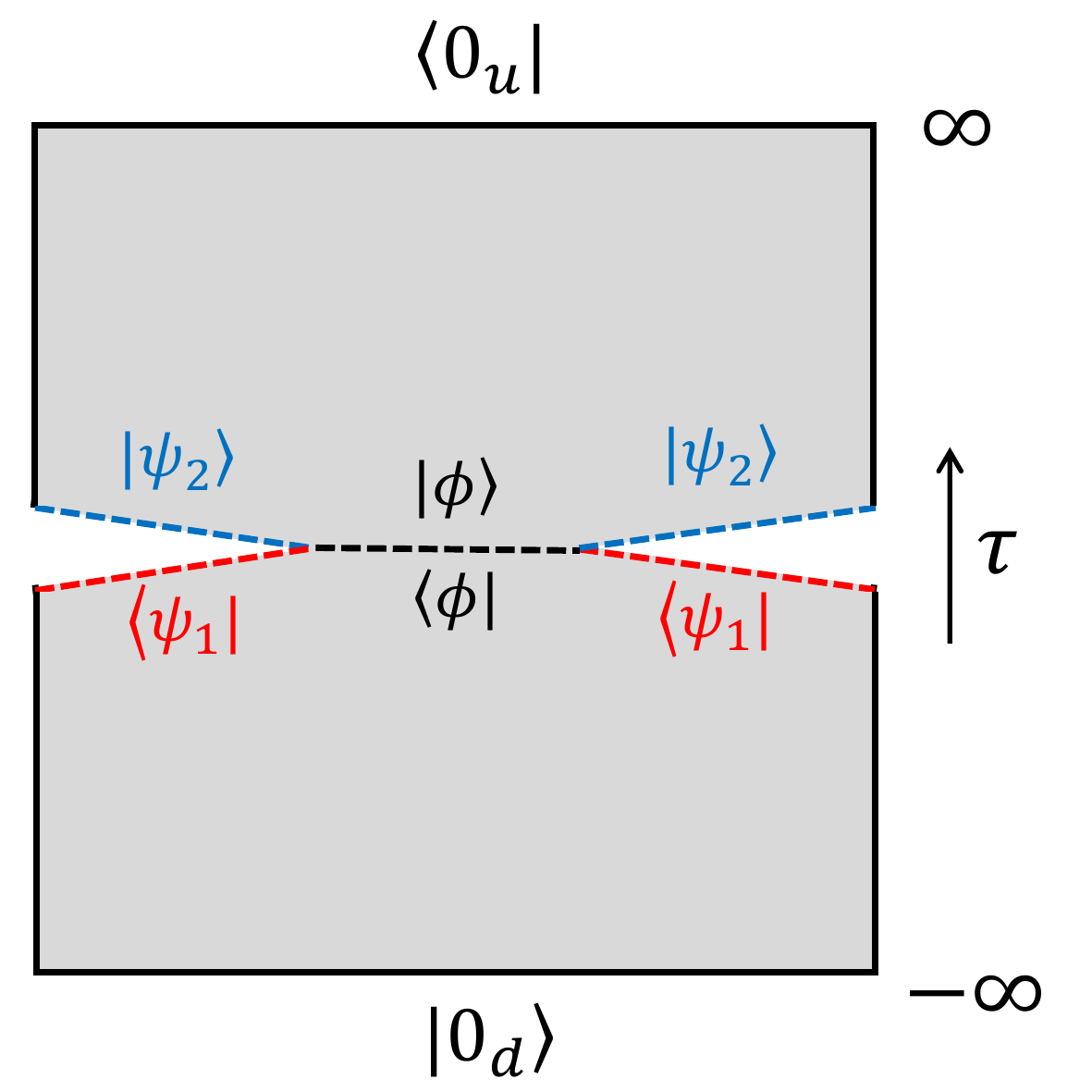}
\end{matrix}\,.\label{TMDvac}
\eea
And, the purity of the TMD density matrix can be further expressed as
\bea
\text{tr}[\rho^2_{\text{TMD}}] 
&=&\sum_{\phi,\bar \phi}\sum_{\psi_1,\psi_2}\bra{\psi_1}\brat{\phi}{0_d}\brat{0_u}{\phi}\ket{\psi_2}
\left[\bra{\psi_1}{\langle {\bar \phi}|{0_d}\rangle}{\langle{\bar\phi}|{0_u} \rangle} \ket{\psi_2}\right]^*\nn\\
&=& \bra{0_u}^*\brat{0_u}{0_d}\ket{0_d}^*\,.\label{vac0}
\eea
Note that the above purity $\text{tr}[\rho^2_{\text{TMD}}] $ can only see the vacua because the system is evolved from the past infinity to future infinity. The excitation modes can not be seen due to the infinite distance between the two-time slices, which was elaborated in \eqref{infinite} for the disk case.

In order to see the gravitational effects in the bulk, we can imagine the down vacuum from the past infinity evolves to a finite time slice $\tau=-d/2$ and starts to feel the gravity. Then the system evolves to time slice $\tau=d/2$. After that gravity disappears and the system evolves to future infinity. This means that gravity is confined in a region between time slices $\tau=-d/2$ and $\tau=d/2$. Suppose the gravitational Hamiltonian is denoted as $\mathcal{H}$, one need to multiply the purity shown in \eqref{vac0} by
\be
Z_{\text{gravity}}=\bra{0_d} e^{-\int_{-d/2}^{d/2}\mathcal{H} \dd \tau}\ket{0_d}\,.
\ee
Diagrammatically, the purity can be illustrated as gluing two copies of (\ref{TMDvac}) together
\be\label{vac}
\text{tr}[\rho^2_{\text{TMD}}] =\sum_{\psi_1,\psi_2}
\begin{matrix}
\includegraphics[width=3cm]{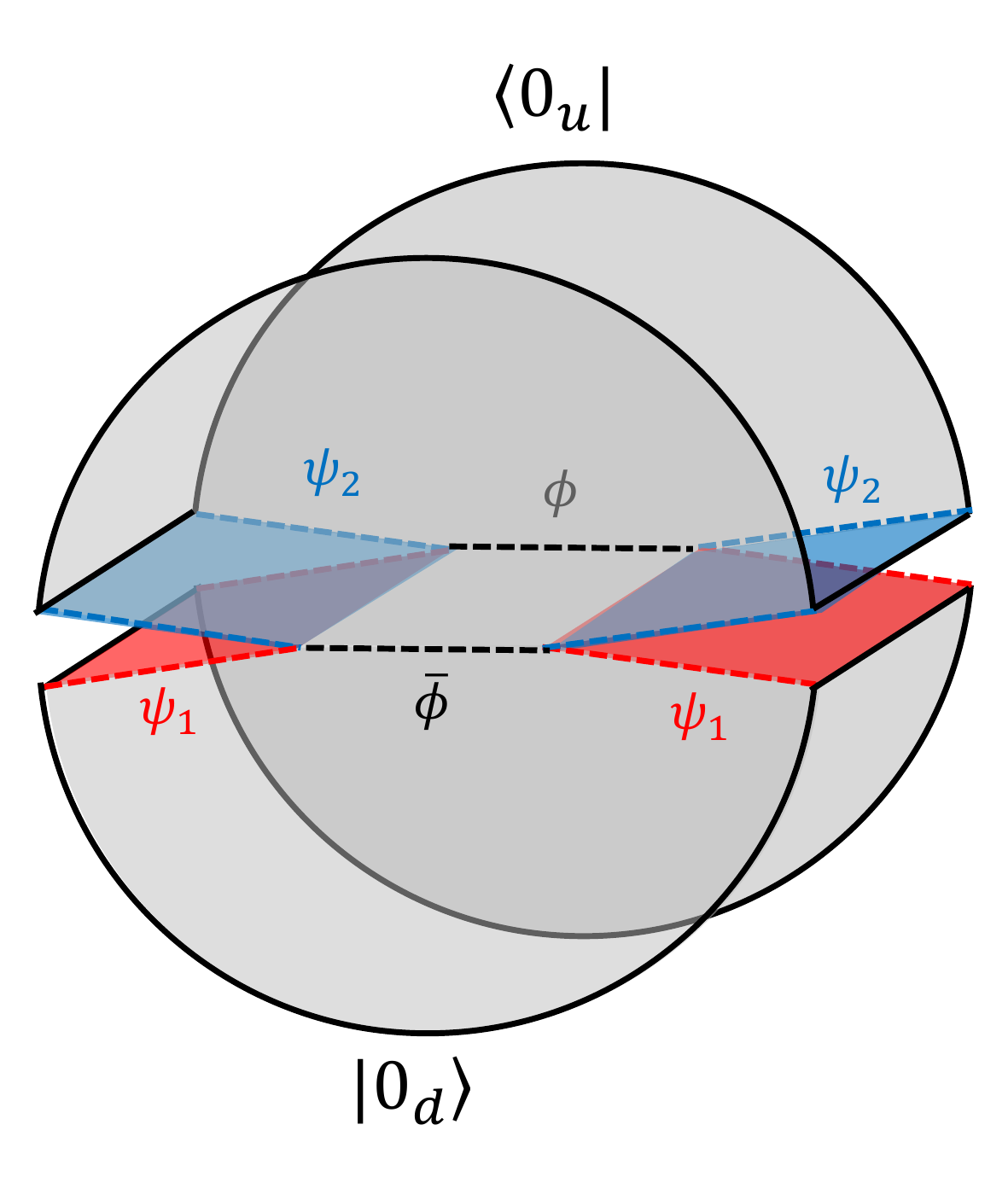}
\end{matrix}
=
\begin{matrix}
\includegraphics[width=2.3cm]{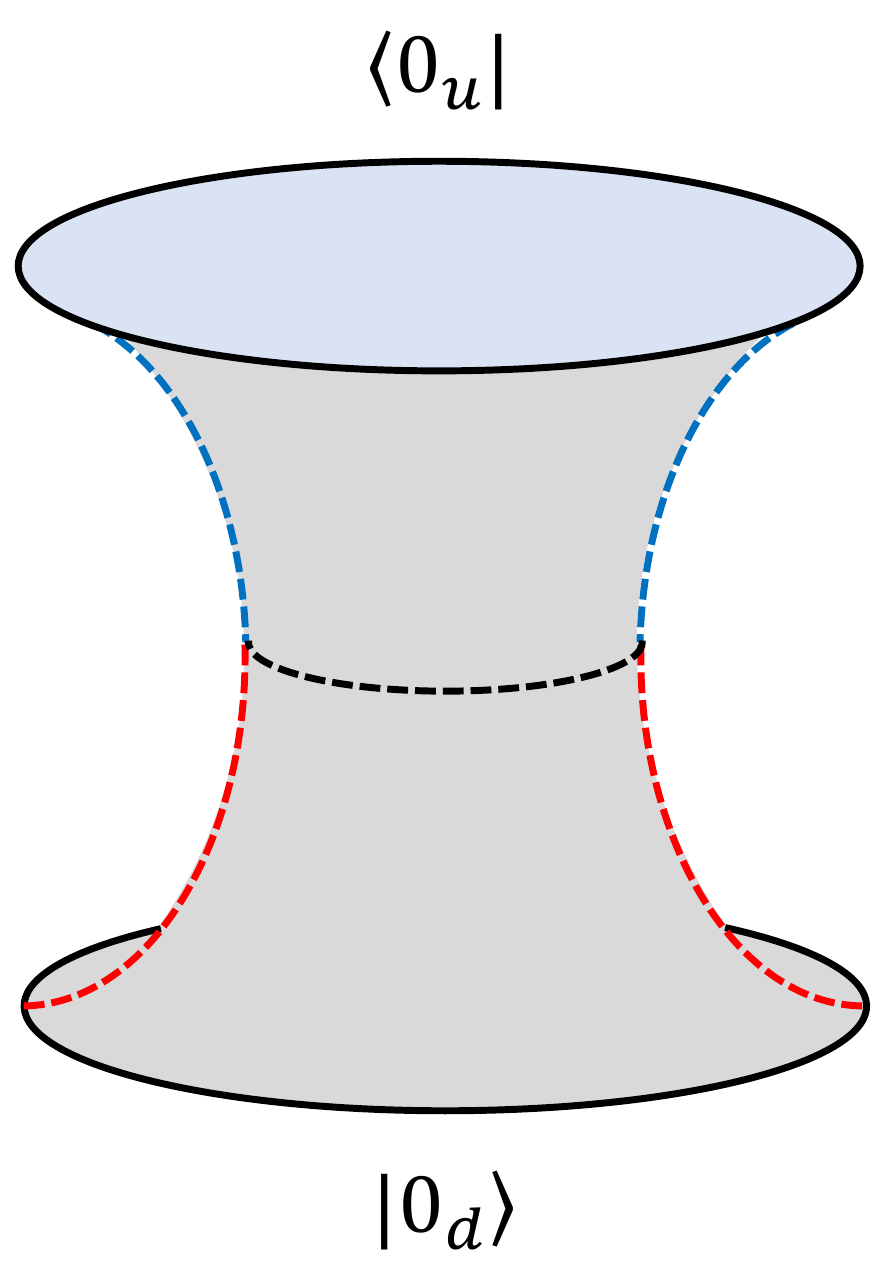}
\end{matrix}\,,
\ee
which can be regarded as the inverse process of (\ref{TMDdensity}).

Comparing equations (\ref{vac0}) and (\ref{vac}), we can conclude that the 2-fold replica wormhole should be proportional to the transition amplitude between the up and down vacua
\be\label{18}
Z_2\propto \bra{0_u}^*\brat{0_u}{0_d}\ket{0_d}^*\times Z_{\text{gravity}}\,.
\ee
Here, we have chosen to project the Hamiltonian onto a specific vacuum. The transition amplitude should always be proportional to the volume of degenerate vacua.
Nevertheless, the point here is that the effect of the transition between different vacua should be included in $Z_2$.
Here we use $\propto$ because the renormalization factor shown in (\ref{replicaA}) is not included.
Now, we can see that the replica wormhole also counts the transition from different vacua evolved from $\tau=-\infty$ to $\tau=\infty$.
It is more clear that, as shown in (\ref{vac}), the state at $\tau=-\infty$ and $\tau=\infty$ are not necessarily the same because of the independent circles, which is different from the Hawking saddle where the up and down circles are connected because of trace operation.
The allowance for different vacua to exist is suggested to understand lots of interesting physics related to wormholes like the factorization puzzle \cite{Blommaert2021,Saad2021,Cheng:2022nra} and the black hole information paradox \cite{Blommaert:2022ucs,Cheng2021a,McNamara2020}.
The conclusion can be easily generalized to $n$-fold replica wormholes, which can be regarded as transitions between more different vacua.

We would like to end this section with some comments on SSB and vacuum degeneracy.
For a complex scalar field theory with Mexican hat potential, the vacuum degeneracy comes from the SSB of the global $U(1)$ symmetry. The minimum of the potential is characterized by a compact parameter $\rho$. Let us suppose the vacuum is chosen at a specific value $\ket{\rho_0}$ and there is a possible generator $Q_\theta$ that acts on the vacuum state and changes the value of $\rho$. The generator can be written as
\be\label{globaltrans}
e^{iQ_\theta}~\ket{\rho_0}\sim \ket{\rho_0+\theta}\,.
\ee
The generator $Q_\theta$ moves the QFT from one vacuum to a new one. 
However, the operator $Q_\theta$ is not in the Hilbert space because it is not normalizable. To see this physically, we can suppose the original vacuum is $\ket{\rho_0}$, and we want to change the global vacuum to a different one $\ket{\rho_1}$. We can start to change it with a bubble of local degrees of freedom with vacuum $\ket{\rho_1}$. There is a non-zero gradient of the scalar field on the edge, and the energy needed to flip the vacuum is proportional to the bubble area.
 Then it needs infinite energy to move all the local degrees of freedom from vacuum $\ket{\rho_0}$ to vacuum $\ket{\rho_1}$. So in QFT, superselection sectors are formed, and a specific vacuum is chosen.

However, there are two possible exceptions, where superselection sectors are not formed.
The first one is low-dimension cases. In two dimensions, the area of the bubble is a point, and no superselection sector is formed. That is why we can twist left and right single trumpets for wormhole geometries in 2-dimensional JT gravity.
The second one is vacuum degeneracy due to the SSB of the large gauge (or diffeomorphism) symmetry.
As opposed to the global case, the operators that move the theory between different vacua are not non-renormalizable operators. This is because would-be gauge symmetries are labeled by local functions, the SSB of which does not need to occur simultaneously everywhere.

As shown by HPS, there are infinite degenerate vacua at null boundaries because of the SSB of large gauge symmetry, and no superselection sector is formed.
We can relate the vacuum degeneracy with the wormhole geometry, and the two-fold wormhole can be thought of as a transition between different soft-hair-generated vacua. 
The replica wormhole can be regarded as the process comparing the would-be gauge parameter at the two boundaries, thus can decrease the entropy of the black hole and radiation.

\section{2-dimensional JT gravity}
\label{JT}

Let us check if the above proposal makes sense in the 2-dimensional Euclidean JT gravity. In this calculable low-dimensional quantum gravity theory, we can check the property of the wormhole geometry.

The action of JT gravity can be written as
\bea
I_{JT}[g,\Phi]&=&-S_0 \chi -\frac{1}{16\pi G_N}\int_{\mM}\sqrt{|g|}\Phi(R+2)\nn\\
&~& 
-\frac{1}{8\pi G_N}
\oint_{\pd \mM}\sqrt{|g|}\Phi(K-1)\,,
\eea
with constant $S_0$ and Euler characteristic $\chi$. With boundary condition 
\be
h_{tt}\big{|}_{\pd\mM}=\frac{1}{\varepsilon^2},~~~~\Phi\big{|}_{\pd\mM}=\frac{\Phi_b}{\varepsilon}\,,
\ee
the boundary action can be written as
\be
I_{bdy}[F]=-C\int\dd \tau \{F, \tau\},~~~~C=\frac{\Phi_b}{16\pi G_N}\,.
\ee
We have $\varepsilon\to 0$ near the boundary, and $\{F, \tau\}$ is the Schwarzian derivative. In the black hole frame, we have 
\be
F(\tau)=\tan \frac{\pi}{\beta}f{\tau}\,.
\ee
The new variable $f(\tau)$ can be interpreted as a time repara-metrization of the boundary thermal circle, with properties
\be
f(\tau+\beta)=f(\tau)+\beta\,,~~~~~f'(\tau)\geq 0\,.
\ee
The bulk geometry is fixed to $R+2=0$, so the boundary theory and $f(\tau)$ are the main object people usually care about in boundary theory.
The boundary partition function, or the bulk disk partition function, can be calculated perturbatively. It was shown by Stanford and Witten \cite{Stanford:2017thb} that the Schwarzian theory is one-loop exact using fermionic localization. The bulk black hole partition function can be written as (see \cite{Mertens:2022irh} for review)
\be
Z_{1}(\beta)=\frac{1}{4\pi^2}\left(\frac{2\pi C}{\beta}\right)^{3/2}e^{S_0+\frac{2\pi^2C}{\beta}}\,.
\ee

Let us consider gravitational partition function with $n$ boundaries
\be
Z(\beta_1,\beta_2,\cdots,\beta_n)\,.
\ee
The partition function includes all possible topologies of the spacetime, and can be decomposed by the Euler characteristic $\chi$, i.e.
\be
Z(\beta_1,\beta_2,\cdots,\beta_n)=\sum_{g}e^{S_0(2-2g-n)}\times Z_{g,n}(\beta_1,\beta_2,\cdots,\beta_n)\,.
\ee
$Z_{0,2}(\beta_1,\beta_2)$ is what we care about in this paper, so let us see what this partition function is in JT gravity. We end up with two parts: the disconnected and the connected geometries. 
For disconnected disks, the partition function is 
\be
Z_1(\beta_1)Z_1(\beta_2)=
\begin{matrix}
\rotatebox{90}{\includegraphics[width=2cm, angle=90]{BH2}}
\end{matrix}
\times
\begin{matrix}
\includegraphics[width=2cm]{BH2}
\end{matrix}\,.
\ee
For connected geometry, we can glue two disk geometries with insertions of defects that break $PSL(2,R)$ to $U(1)$, which are called single trumpets. The partition function of the single trumpet is also one-loop exact and can be evaluated as
\be\label{trumpet}
Z_{\text{trumpet}}(\beta)=\frac{1}{2\pi}\sqrt{\frac{2\pi C}{\beta}}e^{-\frac{C}{2\beta}b^2}\,.
\ee
The remaining $U(1)$ is characterized by the twist parameter $\tau$, and $b$ is the length of the geodesic at the middle of the geometry
\be\label{twsit}
\begin{matrix}
\includegraphics[width=3.3cm]{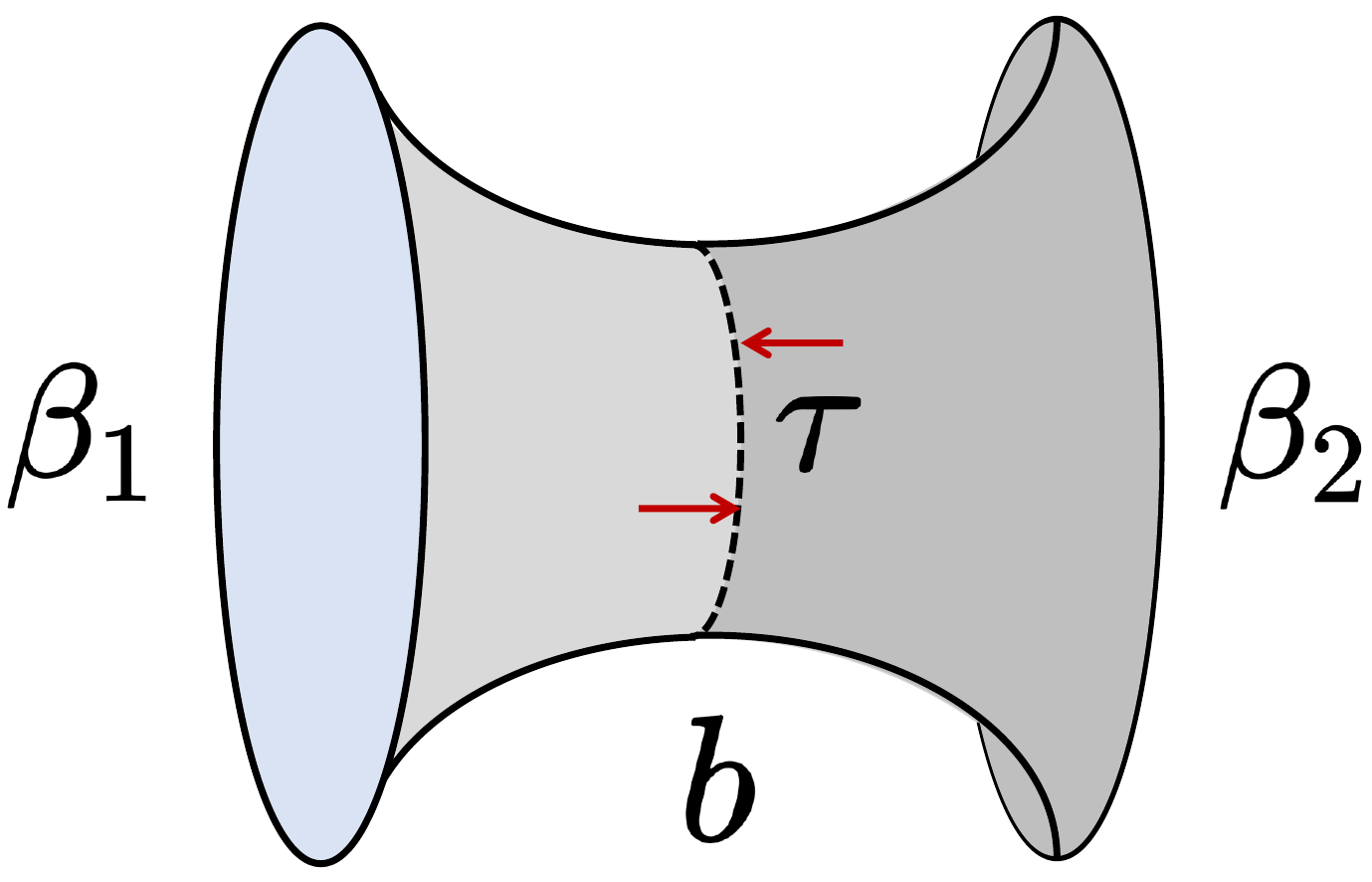}
\end{matrix}\,.
\ee
The wormhole partition function should be an integral over the twist parameter $\tau$, with $\int_0^{b}\dd \tau =b$. This is because \eqref{trumpet} is independent of $\tau$. Then, one can integrate all possible $b$ to count all of the on-shell effects\footnote{The measure can be derived from the first-order formalism \cite{Saad2019}}
\be
Z_{0,2}(\beta_1,\beta_2)=\int_0^{\infty} Z_{\text{trumpet}}(\beta_1)Z_{\text{trumpet}}(\beta_2)~b\dd b\,.
\ee 

Note that we expect the wormhole geometry as a saddle of the path integral, while here the integral is over all possible quantum configurations. Is there a ``stable" wormhole saddle in the above integral? From \eqref{trumpet}, because $b$ is pushed to zero in the exponential. But if we include the twist parameter and extremize the integrand, there is a stable value of $b$\footnote{This is similar to constrained instantons \cite{Cotler:2022rud,Cotler:2020lxj}}.
Solving 
\be
\pd_b[Z_{\text{trumpet}}(\beta_1)Z_{\text{trumpet}}(\beta_2)~b]=0\,,
\ee
one finds the solution
\be
\bar b=\frac{\sqrt{\beta_1\beta_2}}{\sqrt{C(\beta_1+\beta_2)}}\,.
\ee
The corresponding partition function can be written as
\be
\bar Z_{0,2}(\beta_1,\beta_2)=\bar b\times \frac{C}{2\pi \sqrt{e\beta_1\beta_2}}\,.
\ee
$\bar b$ is the length of the twist parameter $\tau$, that labels the $U(1)$ circle, as discussed in this section. 

So, the JT calculation can be regarded as evidence for what we discussed in the previous section, given in \eqref{18}. The wormhole partition function is proportional to the number of vacua generated by the symmetry of the geometry.

\section{Conclusion and discussion}
\label{con}

In this paper, we address the question of how to understand the replica wormhole.
When deriving the island rule from the Euclidean path integral, the replica wormhole saddle naturally arises and plays a vital role in the post-Page time. 
The replica wormhole connects different replicas in the replica trick is a confusing object, and it is important to better understand the physical meaning of these wormhole geometries. 

We have shown that the replica wormhole should be regarded as a transition between different degenerate vacua. More specifically, all the possible vacua in the Hawking saddle are identified because of the trace operation shown in Fig. \ref{replicahawking}, but the situation is different for the replica wormhole saddle in Fig. \ref{replicaworm}. There can be different vacua represented by the left and right black circles, and the purity of the TMD density matrix is shown to be proportional to the transition between two different vacua, which is illustrated in Eq. \ref{vac0} and \ref{vac}. The same argument can be generalized to $n$-fold wormholes.

Therefore, the contribution of the replica wormhole saddles should be controlled by the manifold of the degenerate vacua.
We have checked the above proposal in 2-dimensional JT gravity and explicitly shown that the partition function of the wormhole saddle is proportional to the manifold of the twist parameter. 
Furthermore, following \cite{Pasterski2020,Cheng2020,Cheng2021a}, the wormhole geometries connecting different vacua are suggested to be regarded as the measurement process that compares different vacuum configurations.
The measurement process connecting different vacuum configurations decreases the fine-grained entropy of the system, and the Page curve can also be derived by comparing the rate of the Hawking radiation and measurement processes.

For further studies, we would like to explore a possible connection between the island prescription and the soft BHIP approach, which is related to the attempts to understand the BHIP using the concept of projecting black holes onto specific soft hair configurations \cite{Pasterski2020,Cheng2020,Cheng2021a}. Especially in our recent work \cite{Cheng2020}, we have started to study this possible relation between different wormhole geometries and soft hair measurements, and the disconnected Page curve shown in Fig. 7 of \cite{Cheng2020} can be regarded as transitions between different wormhole saddles. Further connections between the soft hair approach and the replica wormholes might require to look closely at the black hole soft $\mS$-matrix \cite{Gaddam2020,Himwich2020}, which needs further study. In this triumph, the whole island prescription might be eventually understood in terms of HPS's soft hair.

\paragraph{Acknowledgements} We would like to thank Pujian Mao for inspirational discussions. 
P.C. is supported by the National Natural Science Foundation of China (NSFC) under Grant No. 11905156 and 11935009. 
Y.A. is supported by the NSFC under Grant No. 12275238, 11975203, and 11675143.

\appendix
\section{A review of Euclidean path integral}
\label{EPI}

In this appendix, let us review and emphasize some useful concepts in QFT.
This section will discuss the path integral representation of QFT and
some useful concepts in black hole thermodynamics.
All the selected concepts and related interpretations will be crucial for understanding the replica wormhole later.

For a scalar field theory, the transition amplitude between $\ket{\phi_1}$ and $\ket{\phi_2}$ can be represented as the following path integral
\be
\brat{\phi_2(t_2)}{\phi_1(t_1)}=\bra{\phi_2}e^{-iH(t_2-t_1)}\ket{\phi_1}=\int_{\phi(t_1)=\phi_1}^{\phi(t_2)=\phi_2}\mD \phi~ e^{iS}\,.
\ee
Writing everything with Euclidean time $\tau=it$, we can express the transition amplitude as
\be
\brat{\phi_2(\tau)}{\phi_1(0)}=\bra{\phi_2}e^{-\tau H}\ket{\phi_1}=\int_{\phi(0)=\phi_1}^{\phi(\tau)=\phi_2}\mD \phi~ e^{-S_E}\,,
\ee
with the Euclidean action $S_E$. Now we have a Euclidean path integral from $\phi_1$ to $\phi_2$, which can be illustrated diagrammatically as follows
\be
\int_{\phi(0)=\phi_1}^{\phi(\tau)=\phi_2}\mD \phi~ e^{-S_E} =
\begin{matrix}
 \includegraphics[width=2.7cm]{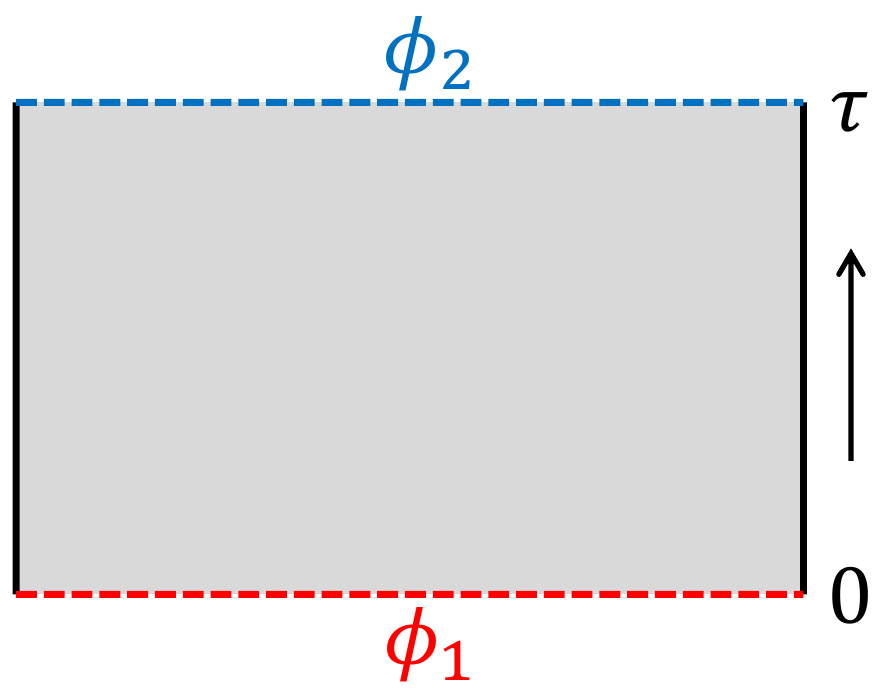}
\end{matrix}\,.
\ee
The path integral is over the grey region, and $\phi_1$ and $\phi_2$ should serve as boundary conditions on a fixed Cauchy surface.
Correspondingly, we can also express the wave function $\ket{\Psi}$ as
\be
\ket{\Psi}=e^{-\tau H}\ket{\phi_1}=\int_{\phi(0)=\phi_1}^{??}\mD \phi~ e^{-S_E} =
\begin{matrix}
 \includegraphics[width=2.5cm]{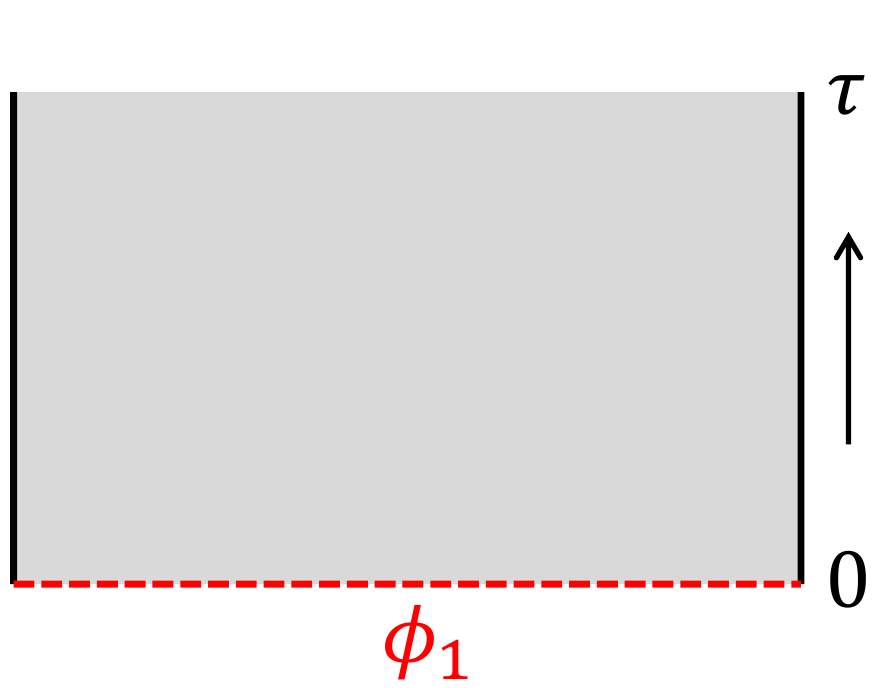}
\end{matrix}\,,\label{waveF}
\ee
\noindent which can be measured by the overlap with state $\bra{\phi_2}$ as $\Psi[\phi_2]=\brat{\phi_2}{\Psi}$. For the wave function shown in (\ref{waveF}), the up boundary condition is not specified.
Following the same logic, the thermal density matrix $\rho$ can be depicted as
\be
\bra{\phi_2}\rho\ket{\phi_1}=\bra{\phi_2}e^{-\beta H}\ket{\phi_1} =
\begin{matrix}
 \includegraphics[width=2.7cm]{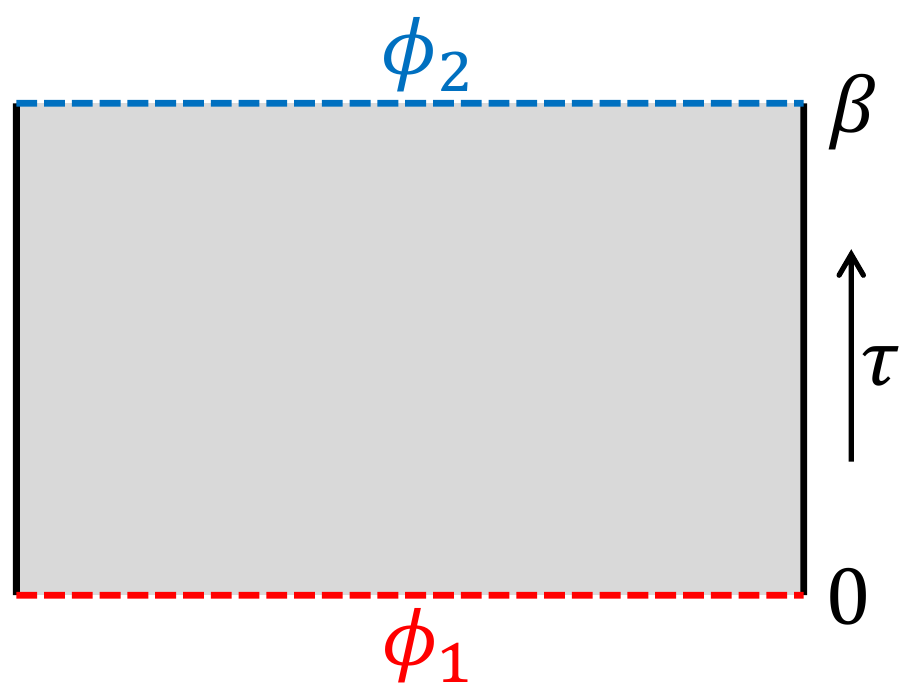}
\end{matrix}\,.
\ee
Taking the trace of the above density matrix gives out the thermal partition function $Z(\beta)$, which can be represented as
\be
Z(\beta)=\text{tr}~e^{-\beta H} =\sum_\phi \braket{\phi}{e^{-\beta H}}=
\begin{matrix}
 \includegraphics[width=2.8cm]{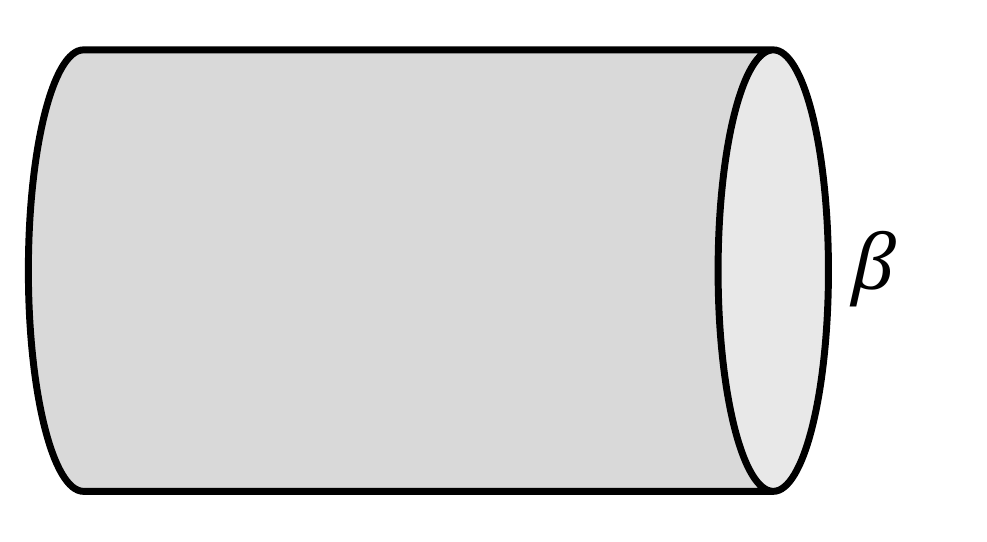}
\end{matrix}\,.\label{PF}
\ee
One can also insert some operators in the bulk to calculate correlation functions, and then  \eqref{PF} can be used to explain why thermal Green's function has periodicity $\beta$.

The ground state $\ket{0}$ can be got by evolving any state $\ket{X}$ for $\infty$ amount of time. Decomposing the state $\ket{X}$ on energy eigenstates $\ket{n}$ with $H\ket{n}=E_n\ket{n}$,
the ground state can be obtained by
\be\label{infinite}
\lim_{\tau\to \infty}e^{-\tau H}\ket{X}=\ket{0}+\lim_{\tau\to \infty}\sum_{n\neq 0} e^{-\tau E_n} \ket{n}\to \ket{0}\,,
\ee
because all the high energy eigenstates with non-zero $E_n$ are largely suppressed by $e^{-\tau E_n}$ as $\tau\to \infty$.
The above argument can be interpreted as, after $\infty$-Euclidean time, all the high energy states die off, and only the vacuum state survives.
Diagrammatically, we can depict the vacuum wave functional as
\be
\brat{\phi_2}{0}=\int_{\phi(-\infty)}^{\phi(0)=\phi_2}\mD \phi~ e^{-S_E} =
\begin{matrix}
 \includegraphics[width=2.8cm]{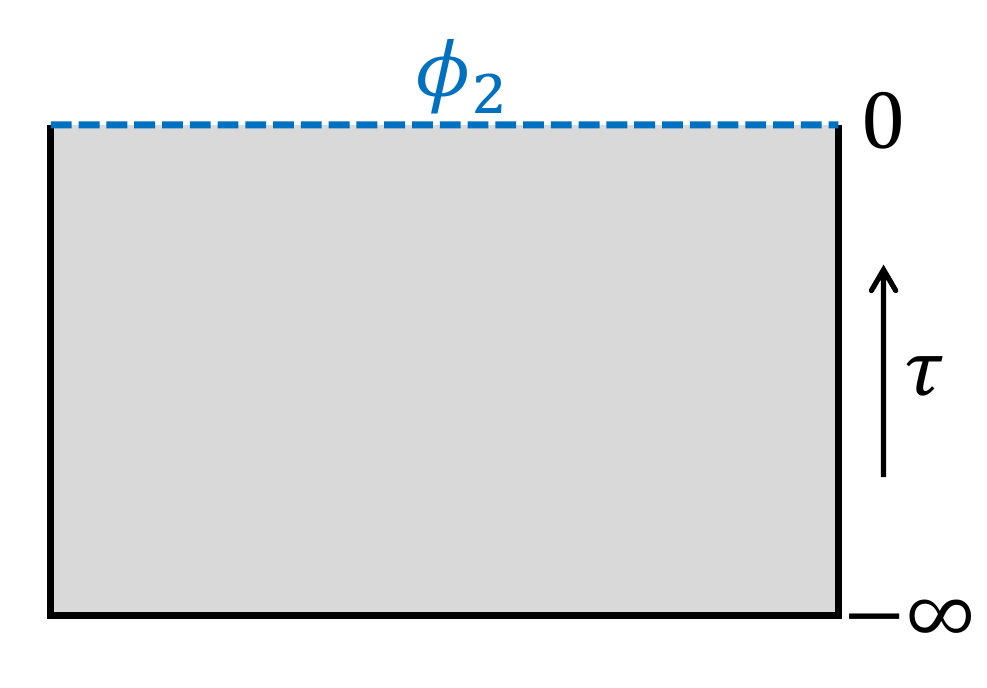}
\end{matrix}\,,\label{vacuum}
\ee
where the solid line labeled by $-\infty$ can be used to represent the vacuum state because only the ground state can survive after the evolution. Furthermore, it is straightforward to see that the vacuum-to-vacuum amplitude $\brat{0}{0}$ can be obtained by inserting an identity
\be
\brat{0}{0}=\sum_\phi \brat{0}{\phi}\brat{\phi}{0}\,,
\ee
whose Euclidean path integral representation can be illustrated as
\be
\brat{0}{0}=\int_{\phi(-\infty)}^{\phi(\infty)}\mD \phi~ e^{-S_E} =
\begin{matrix}
\includegraphics[width=2.8cm]{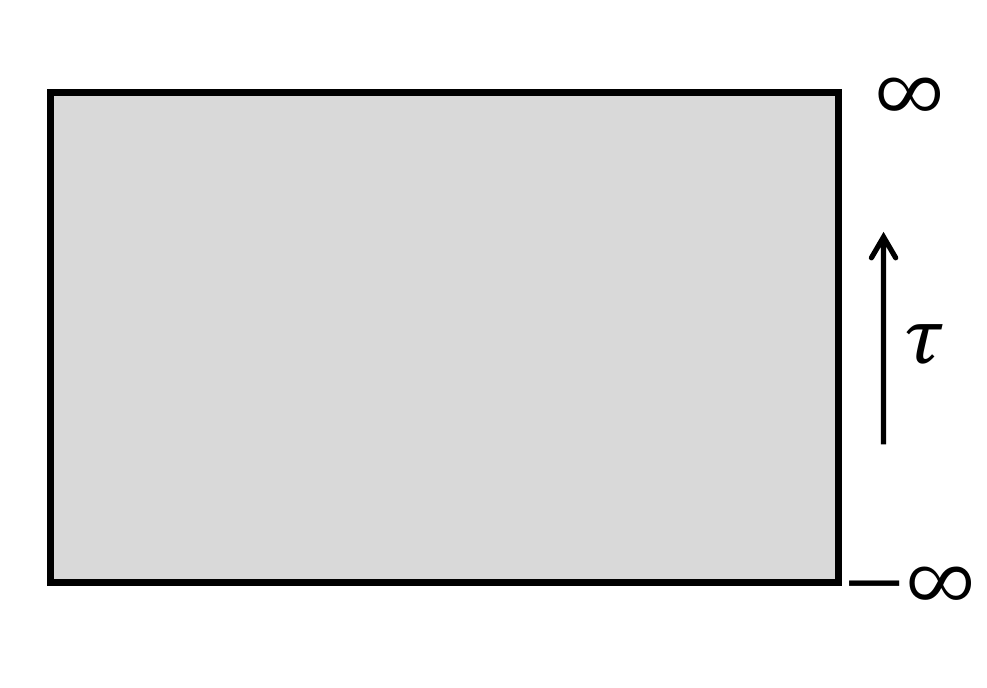}
\end{matrix}\,,
\ee
where we simply glue two diagrams shown in (\ref{vacuum}) together.
All the above diagrams can also be generalized to situations where $\phi$s are defined on boundaries with different topologies, and there can also be handles and holes in the bulk.

%

Next, let us consider the Rindler space 
and derive the Unruh effect from the Minkowski vacuum. The density matrix of the Minkowski vacuum can be written as
\be
\rho=\ket{0}_M\bra{0}\,.
\ee
Because of the presence of the Rindler horizon, we divide the spatial surface into two parts, the $r>0$ region $A$ and the $r<0$ region $B$. The reduced density matrix in region $A$ can be written as
\be
\rho_A=\text{tr}_B \ket{0}_M\bra{0}\,,
\ee
whose path integral representation is
\bea
\bra{\phi_2}\rho_A\ket{\phi_1}=\sum_{\phi_B}\bra{\phi_2}\otimes \brat{\phi_B}{0}_M\brat{0}{\phi_B}  \otimes \ket{\phi_1}=
\begin{matrix}
 \includegraphics[width=2.5cm]{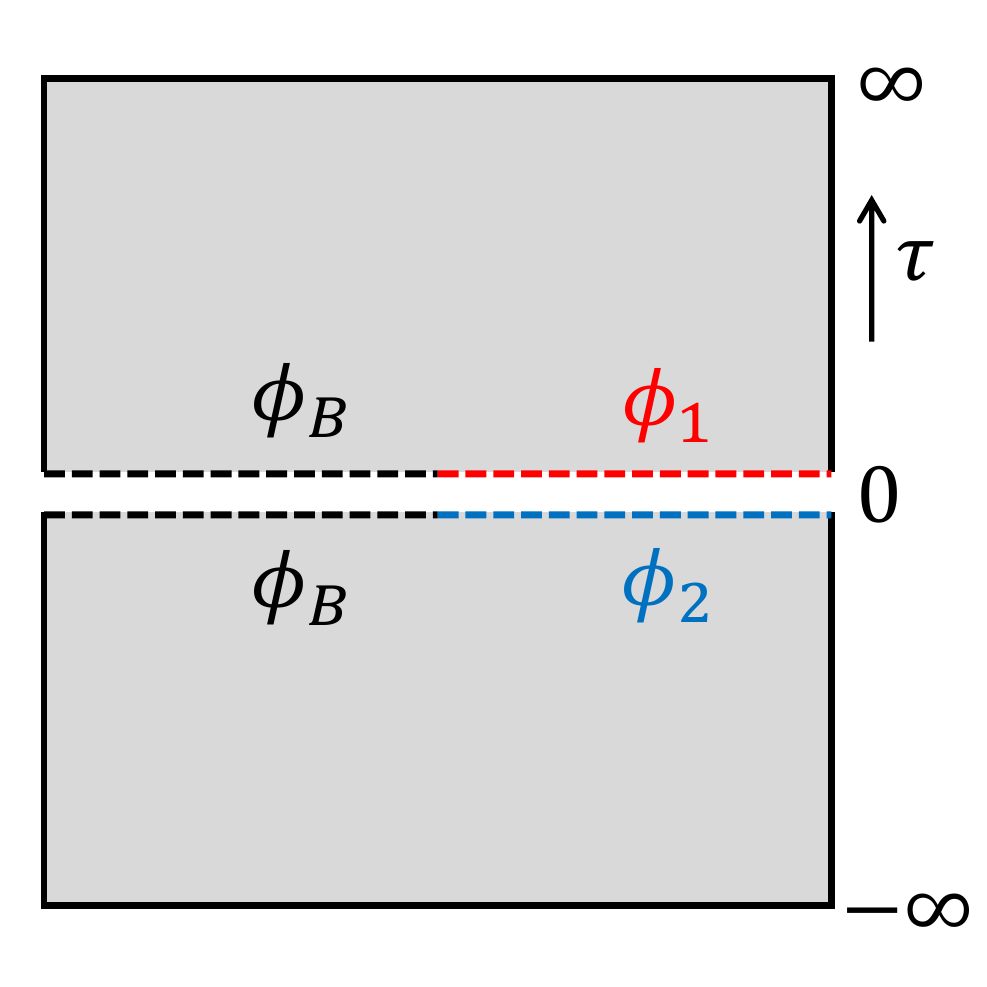}
\end{matrix}\,.
\eea
The trace over fields $\phi_B$ in region $B$ glues the $B$ region together. With a different foliation, the above density matrix can be depicted as
\be
\sum_{\phi_B}
\begin{matrix}
 \includegraphics[width=2.2cm]{density1}
\end{matrix} =
\begin{matrix}
 \includegraphics[width=2.2cm]{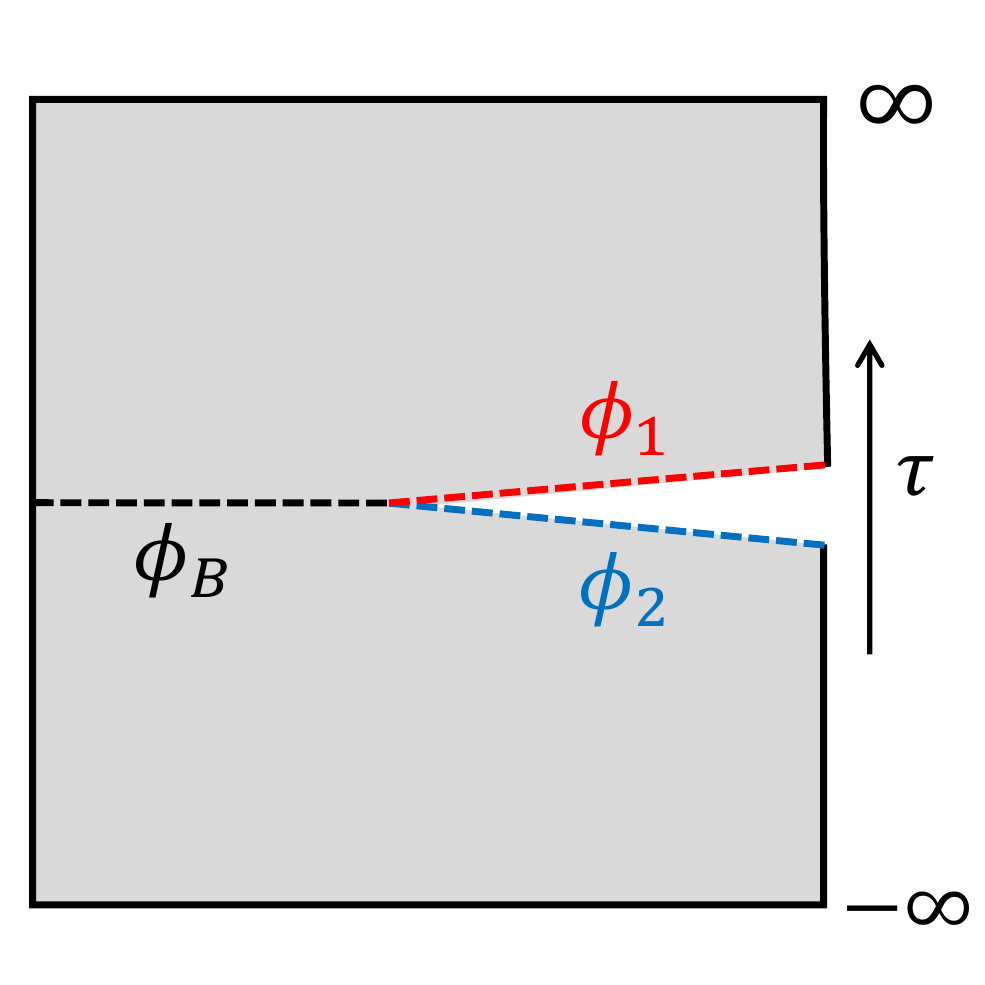}
\end{matrix} =
\begin{matrix}
 \includegraphics[width=2.2cm]{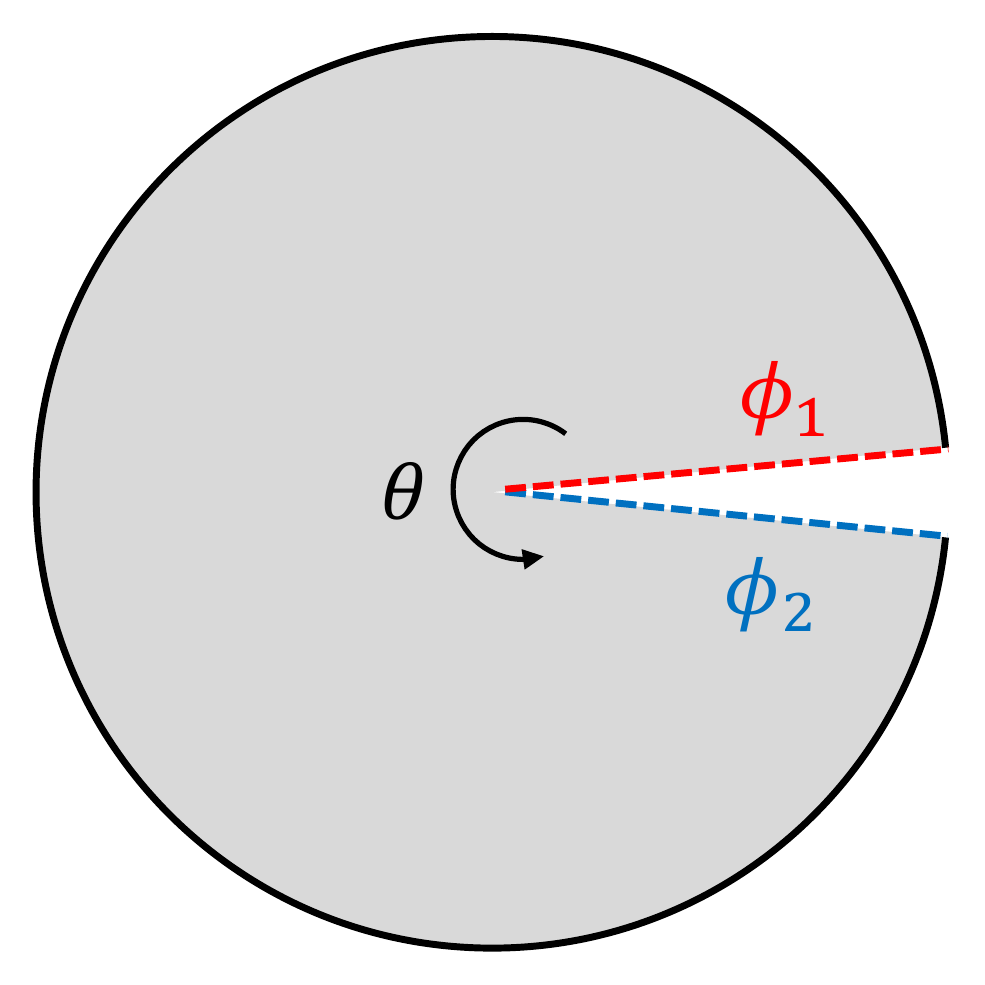}
\end{matrix}\,, \label{pacman}
\ee
where the $\theta$ direction can be regarded as the Euclidean time for Rindler spacetime. So the reduced density matrix $\rho_A$
can be regarded as a thermal density matrix with inverse temperature $\beta=2\pi$, which can be written as
\be
\bra{\phi_2}\rho_A\ket{\phi_1}=\bra{\phi_2}e^{-2\pi H_{\theta}}\ket{\phi_1}\,,
\ee
with the Rindler Hamiltonian $H_{\theta}$. Now, we can identify the inverse temperature of Rindler space as $\beta=2\pi$. Let us call the figure shown in equation (\ref{pacman}) ``Pacman". Now, we have derived the Unruh density matrix by tracing out the $B$ region of the Minkowski vacuum.
Besides the density matrix, the Minkowski vacuum itself can be treated from a different point of view.
The vacuum wave functional measured by the overlap with state $\bra{\phi_2}\otimes \bra{\phi_1}$ can be represented as
\be\label{TFD1}
\bra{\phi_2}\otimes \brat{\phi_1}{0}_M=
\begin{matrix}
\includegraphics[width=2.5cm]{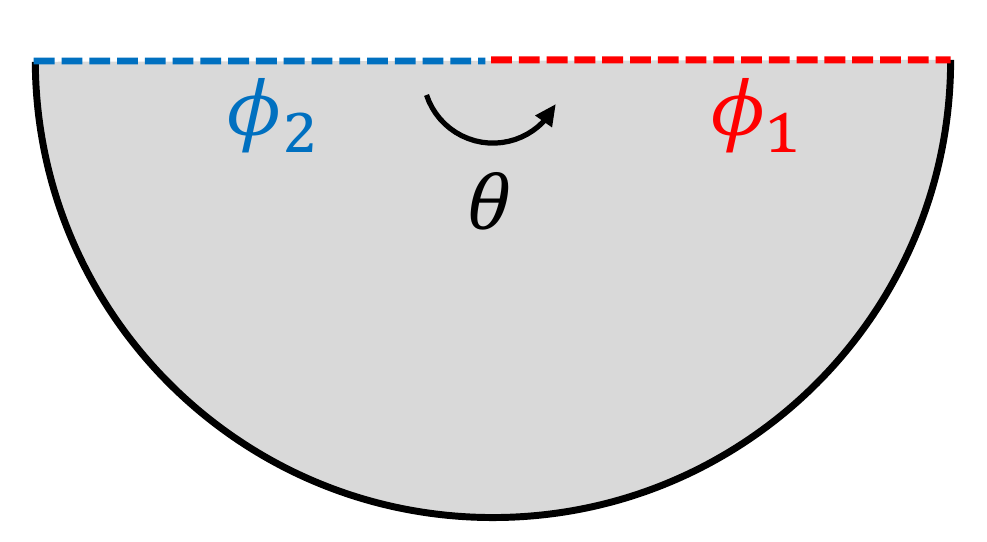}
\end{matrix}=
\bra{\phi_1}e^{-\frac{\beta}{2} H_{\theta}}\ket{\phi_2}\,,
\ee
which can be further expressed as
\be\label{TFD2}
\bra{\phi_1}e^{-\frac{\beta}{2} H_{\theta}}\ket{\phi_2}=\sum_n e^{-\frac{\beta}{2} E_n}\brat{\phi_1}{n_1}\otimes\brat{\phi_2}{n_2}^*\,.
\ee
$\ket{n}$ is the eigenstates of Rindler Hamiltonian $H_{\theta}$ on the left and right sides, with $H_{\theta}\ket{n}=E_n\ket{n}$. The * symbol represents the CPT transformation. Comparing (\ref{TFD1}) and (\ref{TFD2}), we can conclude that the Minkowski vacuum can be regarded as a purification of the Rindler Hilbert space
\be
\ket{0}_M=\sum_n e^{-\frac{\beta}{2}H_{\theta}}\ket{n_2}^*\otimes\ket{n_1}\,,
\ee
which is called thermo-field double (TFD) state \cite{Maldacena2001}. We have shown the origin of the thermal spectrum for the Rindler observer restricted to the Rindler patch. The same argument can be generalized to the black hole case, which can be used to explain the thermal spectrum of the Hartle-Hawking state. The so-called ``cigar" geometry replaces the disk partition function obtained from tracing the Pacman figure because of the curvature of the black hole. The black hole partition function can be represented diagrammatically as
\be
Z_{BH}=\sum_{\phi}
\begin{matrix}
 \includegraphics[width=2.3cm]{BH1}
\end{matrix}=
\begin{matrix}
 \includegraphics[width=2.3cm]{BH2}
\end{matrix}\,.
\ee


\newpage
\providecommand{\href}[2]{#2}\begingroup\raggedright\endgroup

\end{document}